\documentclass[aps,prb,twocolumn,
    groupedaddress,superscriptaddress,
    amsfonts,amssymb,amsmath,floatfix,
    citeautoscript]{revtex4-1}
\usepackage{graphicx} 
\usepackage[centering,hmargin=18mm,tmargin=28mm,bmargin=28mm]{geometry}
\usepackage{amsmath}
\usepackage{hyperref} 
\usepackage{multirow}
\usepackage{newtxtext}                                        
\usepackage{booktabs}
\usepackage{xcolor} 
\usepackage{float}
\usepackage[normalem]{ulem}
\usepackage[textwidth=1.5cm,textsize=footnotesize]{todonotes}
\hypersetup{colorlinks, 
    linkcolor={blue!75!black!80!yellow},
    citecolor={blue!75!black!80!yellow}, 
    urlcolor={blue!75!black!80!yellow}
    }
\usepackage[cmintegrals]{newtxmath}

\makeatletter
\renewcommand\@make@capt@title[2]{%
    \@ifx@empty\float@link{\@firstofone}{\expandafter\href\expandafter{\float@link}}%
    \sffamily{\textbf{#1}}\@caption@fignum@sep#2
}%

\makeatother

\usepackage{graphicx}
\usepackage{mathtools}
\usepackage{braket}
\usepackage{hyperref}
\usepackage{color}
\usepackage{gensymb}

\begin{document}

\title{Excited-State Nanophotonic and Polaritonic Chemistry\\ with \emph{Ab initio} Potential-Energy Surfaces} 

\author{Johannes Flick}
\email[Electronic address:\;]{flick@seas.harvard.edu}
\affiliation{John A. Paulson School of Engineering and Applied Sciences, Harvard
University, Cambridge, Massachusetts 02138, USA}
\author{Prineha Narang}
\email[Electronic address:\;]{prineha@seas.harvard.edu}
\affiliation{John A. Paulson School of Engineering and Applied Sciences, Harvard
University, Cambridge, Massachusetts 02138, USA}

\date{\today}

\begin{abstract}
Advances in nanophotonics, quantum optics, and low-dimensional materials have enabled precise control of light-matter interactions down to the nanoscale. Combining concepts from each of these fields, there is now an opportunity to create and manipulate photonic matter via strong coupling of molecules to the electromagnetic field. Towards this goal, here we introduce a first principles framework to calculate polaritonic excited-state potential-energy surfaces for strongly coupled light-matter systems. In particular, we demonstrate the applicability of our methodology by calculating the polaritonic excited-state manifold of a Formaldehyde molecule strongly coupled to an optical cavity. This proof-of-concept calculation shows how strong coupling can be exploited to alter photochemical reaction pathways by influencing avoided crossings. Therefore, by introducing an \emph{ab initio} method to calculate excited-state potential-energy surfaces, our work opens a new avenue for the field of polaritonic chemistry.
\end{abstract}

\date{\today}

\maketitle

Interest in strong light-matter coupling, at the interface of chemistry, nanophotonics and quantum optics, has accelerated with experiments that explore a new regime of polaritonic chemistry~\cite{ebbesen2016}, where matter and the electromagnetic degrees of freedom become equally important. In photochemistry, the idea of altering molecular processes by strongly coupling to the vacuum fluctuations of the electromagnetic field has been suggested and now demonstrated in various experiments. These examples include the modification of chemical processes in optical cavities under electronic strong coupling~\cite{hutchison2012, stranius2018, Peyskens:2018fk}, as well as strong vibrational coupling~\cite{george2016, thomas2019}. Other prominent systems that have been explored are nanoplasmonic systems in the single (few) molecule strong-coupling regime~\cite{benz2016, kazuma2018, ojambati2019}, few-layer quantum materials~\cite{kleemann2017, basov2017towards}, surface plasmon polaritons~\cite{memmi2017, christensen2017quantum, NatComSundararaman2014}, plasmon-exciton systems~\cite{bisht2019}, and organic systems~\cite{coles2014, munkhbat2018}. These exciting new developments necessitate theoretical and computational methods that explicitly describe both the matter (electronic and nuclear) degrees of freedom, as well as the degrees of freedom of the electromagnetic field.\cite{flick2018c} 

Theoretically, the effect of strong light-matter coupling has been analyzed for photochemical reactions~\cite{galego2016, feist2017b, mandal2019}, thermally-activated chemical reactions~\cite{campos2019}, cavity femtochemistry~\cite{kowalewski2016,kowalewski2016b,bennett2016}, photoisomerization~\cite{fregoni2018}. Additionally, the implications of changes in the ground-state under ultra-strong coupling~\cite{martinez2017, galego2018}, cavity-controlled chemistry via a polaron-decoupling~\cite{herrera2016}, and new spectroscopic observables~\cite{ruggenthaler2017b, flick2018c} have been studied. Traditional \emph{ab initio} methods that solve the electronic-structure problem in quantum chemistry include density-functional theory (DFT)~\cite{kohn1999, ullrich2011}, Hartree-Fock theory~\cite{mclachlan1964}, and coupled cluster methods~\cite{bartlett2007}, among others. However, the use of these methods is limited in the regime of strong light-matter coupling, since these methods describe the electronic interaction, but cannot treat the photonic interactions explicitly and self-consistently. 

To overcome this obstacle, electronic-structure methods have been generalized to include the effects of the matter-photon interaction from first principles. These ideas lead to the development of quantum-electrodynamical density-functional theory (QEDFT)~\cite{tokatly2013,ruggenthaler2014}. More recently a practical exchange-correlation functional was introduced~\cite{pellegrini2015,flick2017c}, and QEDFT extended to account for nuclear motion~\cite{flick2018b}. Related methods from electronic-structure theory that have been extended to the realm of electron-photon interactions are the exact factorization approach~\cite{hoffmann2018, abedi2018}, a conditional decomposition approach~\cite{albareda2019}, and a Maxwell-Hartree-Fock wavefunction approach~\cite{rivera2018}, as well as the extensions of the Born-Oppenheimer approximation~\cite{galego2015, flick2017b, schaefer2018}. However, since many chemical reactions involve excited states of matter, it is of critical importance to correctly describe not only electronic ground states, but also the electronic excited-state manifold. While DFT can be used to calculate ground-state properties, time-dependent density-functional theory (TDDFT)~\cite{runge1984, ullrich2011} can be used to access excited-state properties. TDDFT is often employed in the linear-response regime~\cite{casida1996}. A novel first-principles method based on linear-response theory~\cite{flick2018} to calculate excited-state properties of correlated light-matter systems from first principles was recently formally introduced, but it has yet to be demonstrated in photochemistry.

In this \emph{Letter} we close this critical gap in \emph{ab initio} excited-state methods and study how the excited state manifold of a real molecule is altered under strong light-matter coupling. We analyze the Formaldehyde molecule ($\text{H}_2\text{CO}$) strongly coupled to an optical cavity as a prototype system for a photochemically active system under strong light-matter coupling. Specifically, we construct the potential-energy surfaces including the $^1A_1$ excited-state manifold along the $\text{CO}$ stretch vibrational mode and analyze how these surfaces are altered in the strong-coupling regime. Importantly, these surfaces feature avoided crossings that are due to a mixing of different orbital configurations. For the excited-state manifold of the Formaldehyde molecule, our calculations find states that are a mixture of Rydberg states, as well as more localized valence states. Our choice of the Formaldehyde molecule is motivated by the molecule being an excellent model system since its electronic structure has been studied in great detail using various theoretical tools, such as the multireference configuration interaction method~\cite{hachey1995, hachey1996}, time-dependent density-functional theory (TDDFT)~\cite{casida1998}, and coupled cluster theory including single and double excitations~\cite{gomez2010}. In the following, we will introduce the methodology to calculate and analyze the Formaldehyde molecule potential-energy surfaces under strong-light matter coupling. We emphasize the methods presented here are general beyond the model of a Formaldehyde molecule.

There are different ways to define potential-energy surfaces for light-matter systems, described in detail in Refs.~\citenum{flick2017,flick2017b,schaefer2018, galego2015, feist2018}. All of these possibilities are based on the Born-Oppenheimer approximation (BOA)~\cite{born1927, tully2000} which dramatically simplifies the calculation of electron-nuclear systems. There are two possibilities to perform a calculation along the lines of the Born-Oppenheimer approximation: The cavity Born-Oppenheimer approximation, which is advantageous in the vibrational strong-coupling regime~\cite{flick2017} and the theory of polaritonic surfaces~\cite{galego2015,feist2018}, which is primarily invoked in the electronic strong coupling regime. The latter is most relevant to this work where we want to study the effect of strong coupling on photochemical processes, and to this end, throughout this paper, we refer to the scheme outlined in Refs.~\citenum{galego2015, feist2018} to construct polaritonic surfaces. Using this partitioning, we can factorize the electron-nuclear-photon problem into an electron-photon and a nuclear problem. The emerging potential-energy surfaces then consist of electron-photon (polaritonic) energies. In the following, we outline how we can use linear-response QEDFT to calculate correlated electron-photon energies for different nuclear configurations.

To construct potential-energy surfaces within the limit of strong-light matter coupling, we use the previously introduced linear-response formalism of QEDFT described in Ref.~\citenum{flick2018}, which builds on the formalism in linear response DFT. Linear response DFT is usually either performed in the time-domain by an explicit time propagation~\cite{yabana1999}, or in the frequency-domain~\cite{casida1996, casida2012}. Depending on the size of the problem, one or the other formulation is more computationally efficient. In the frequency domain, the equations to be solved can be formulated as a pseudo-eigenvalue problem, also known as the \textit{Casida equation}. Solving this eigenvalue equation yields in principle the exact many-body excitation energies of the system. In practice, usually two approximations are invoked to calculate these energies: \textbf{(1)} The adiabatic approximation, which neglects any frequency dependence in the exchange-correlation kernel; this simplifies the pseudo-eigenvalue equation, which has to be solved iteratively to an eigenvalue equation. \textbf{(2)} The adiabatic exchange-correlation kernel is only used approximately. Typical approximations are based on the local-density approximation~\cite{kohn1965}, generalized-gradient approximations~\cite{perdew1996}, or hybrid functionals that include exact-exchange~\cite{heyd2003}. The linear-response formalism for electronic systems is well-established and has been applied to a variety of systems, see e.g. Refs.~\cite{jamorski1996, casida2000, casida2012, yang2013} and references therein.

In this work we calculate polaritonic potential-energy surfaces in the non-relativistic limit and the length-frame~\cite{ruggenthaler2014}. Here, we describe a system of $N_e$ interacting electrons coupled to $N_p$ photon modes. Each photon mode $\alpha$ is defined by its frequency $\omega_\alpha$ and electron-photon coupling strength $\boldsymbol \lambda_\alpha$~\cite{tokatly2013, ruggenthaler2014}.
The quantity of interest is the basic variable, i.e. the electron density $n(\textbf{r},t)$, and the electric displacement coordinate $q_\alpha(t)$~\cite{ruggenthaler2014}. $q_\alpha(t)$ is directly connected to the electric displacement field~\cite{pellegrini2015}. It has been shown~\cite{tokatly2013, ruggenthaler2014} that there exists an one-to-one correspondence with this set of basic variables ($n(\textbf{r},t)$, $q_\alpha(t)$) and external variables ($v_\text{ext}(\textbf{r},t)$, $j_\alpha(t)$) for a given initial state, where $v_\text{ext}(\textbf{r},t)$ describes an electronic external potential and $j_\alpha(t)$ a time-derivative of an external current acting on the photonic system. Using this one-to-one correspondence, QEDFT linear response can be formulated as a pseudo-eigenvalue equation as follows~\cite{flick2018} :
\begin{align}
\left(
\begin{array}{ c c  }
U &   V     \\
V^{*} &  \omega_{\alpha}^{2}
\end{array}
\right)
\left(
\begin{array}{ c }
\textbf{E}_{1}  \\
\textbf{P}_{1}  
\end{array}	
\right) 
= 
\Omega_{k}^{2}
\left(
\begin{array}{ c }
\textbf{E}_{1} \\
\textbf{P}_{1} 
\end{array}	
\right), \label{eq:casida-equation}
\end{align}
where the matrices $U$, $V$ are given explicitly by:
\begin{align}
U_{ll'} =& \delta_{ll'}\omega_{l}^{2} + 2\sqrt{\omega_{l}\omega_{l'}}K_{ll'}(\Omega_{k}), \label{eq:matrix1}\\
V_{l\alpha} =& 2\sqrt{\omega_{l}M_{\alpha l}(\Omega_{k})N_{\alpha l}\omega_{\alpha}  }     ,\label{eq:matrix2}
\end{align}
The indices $l {= (a,i)}$ describe transitions from the electronic occupied {$(i)$} to unoccupied states {$(a)$}, where the difference in Kohn-Sham energies yields $\omega_l=\epsilon_a-\epsilon_i$, and the corresponding Kohn-Sham orbitals $\varphi_i(\textbf{r})$, $\varphi_a(\textbf{r})$. The dimensionality of the full matrix in Eq.~\ref{eq:matrix2} is given by a sum of the number of occupied-unoccupied pairs $\mathcal{N}_\text{ou}$ that are considered in the calculation and the number of photon modes $N_p$. The eigenvectors of the matrix in Eq.~\ref{eq:casida-equation}, $\textbf{E}_{1}$ and $\textbf{P}_{1}$, yield the oscillator strengths of the coupled matter-photon system~\cite{flick2018}. For the matrix of Eq.~\ref{eq:matrix2}, we find a block-structure, where $U$ describes the electronic block, $\omega_\alpha^2$ the photonic block and $V$ leads to a coupling of electronic and photonic excitations. If the electronic and photonic system are not coupled, i.e. $\boldsymbol \lambda_\alpha = 0$ and $V=0$, we recover the \textit{Casida equation}~\cite{casida1996}, if only $U$ is considered.

The kernel function $K$ in Eq.~\ref{eq:matrix1} that leads to a mixing of different electronic excitations can be calculated by employing the exchange-correlation kernel $f^n_\text{Mxc} = {\delta v_\text{Mxc}}/{\delta n}$ for a given Kohn-Sham potential $v_\text{Mxc}$ and reads explicitly
\begin{align}
K_{ai,jb}(\Omega_q) &= \iint d\textbf{r} d\textbf{r}'\varphi_i(\textbf{r})\varphi_a^*(\textbf{r})  f^n_\text{Mxc}{(\textbf{r},\textbf{r}',\Omega_q)}\varphi_b(\textbf{r}')\varphi^{*}_j(\textbf{r}'). \label{eq:fxc}
\end{align}
In practice, $f^n_\text{Mxc}$ has to be approximated. Using $f^n_\text{Mxc} = f^n_\text{Hxc} +f^n_\text{pxc}$, we can formally divide the exchange-correlation kernel into a part that describes the electron-electron interaction (Hxc) in the system, and a second part that incorporates all effects due to the electron-photon interaction (pxc). 

Eq.~\ref{eq:matrix2} contains the quantities $N$ and $M$ that couple the matter and the photon block. These are given by:
\begin{align}
M_{\alpha,ai}(\Omega_q)  =&\int d\textbf{r} \varphi_i(\textbf{r})\varphi_a^*(\textbf{r}) f^{q_\alpha}_\text{Mxc}{ (\textbf{r},\Omega_q)} , \label{52b} \\
N_{\alpha,ia} =&  \int d\textbf{r} \varphi^*_i(\textbf{r})\varphi_a(\textbf{r}) g^{n_{\alpha}}_\text{M}(\textbf{r})/{2\omega_\alpha^2} . \label{52c}
\end{align}
with the kernel functions
\begin{align*}
f^{q_\alpha}_\text{Mxc}(\textbf{r}t,t') &= \frac{\delta v_\text{Mxc}(\textbf{r}t)}{\delta q_\alpha(t')}, \\
g^{n_\alpha}_\text{M}(t,\textbf{r}'t') &= \frac{\delta j_{\alpha,\text{M}}(t)}{\delta n(\textbf{r}'t')}.
\end{align*}
that can be derived from the Kohn-Sham potential and current $\delta v_\text{Mxc}(\textbf{r}t)$ and $\delta j_{\alpha,\text{M}}(t)$. For its explicit form, we refer to Ref.~\citenum{flick2018}. In this work, we employ the adiabatic approximation and use the pRPA kernel to describe the electron-photon interaction for $f^n_\text{pxc}$~\cite{flick2018}.

For the case of the Hartree exchange-correlation functional (Hxc), we need an accurate description of our system not only at the equilibrium position, but also for more extended nuclear configurations. In the case of the excited-state manifold of the Formaldehyde molecule, we calculate states that are a mixture of Rydberg states, as well as more localized valence states, such as the $\pi$, and $\pi^*$ states~\cite{casida1998}. Many density-functional approximations perform rather well at the equilibrium distance, but fail to correctly describe the asymptotic behavior of the long-range Coulomb interaction. To overcome this limitation, several methods have been developed that yield an asymptotically correct functional~\cite{becke1988, leeuwen1994} or that asymptotically correct exchange-correlation functionals~\cite{andrade2011}. In this work, we use a correction scheme developed by Casida and Salahub~\cite{casida1996, casida2000}. This scheme combines the local-density approximation with the asymptotically correct potential of van Leeuwen and Baerends~\cite{leeuwen1994} in the SCF step. For the kernel in Eq.~\ref{eq:fxc}, we use the local-density approximation~\cite{perdew1992}. This approach has been demonstrated to yield accurate low-lying excited-state energies ~\cite{casida1998a, schipper2000} and agrees with findings in Ref.~\citenum{hofman2018} that show that quality of the Kohn-Sham eigenvalues in the SCF step may be more important than the quality of the the exchange-correlation kernel.

To compute the polaritonic surfaces, we use a locally-modified version of the octopus code~\cite{castro2006, andrade2014} and the implementation used in this work will be made publicly available in future. While we explicitly describe all valence electrons in our simulations on a real-space grid, we use the pseudodojo library~\cite{vansetten2018} to describe the core-electrons with pseudopotentials. We study the potential-energy surfaces of Formaldehyde along the $\text{CO}$-stretch coordinate for frozen $\text{CH}_2$ group. We keep $\text{CH}_2$ at its experimental geometry $\text{R}_\text{CH}=1.10$~\AA,~ and $\text{HCH} = 116.3\degree$ to be consistent with previous calculations in literature~\cite{hachey1995, casida1996}. We choose a large simulation box consisting of spheres with radius $6$~\AA~ around each atom and a grid-spacing of $0.14$~\AA. This grid allows us to equally well capture the long-range Rydberg states, as well as the bound valence states. To diagonalize the matrix in Eq.~\ref{eq:matrix2}, we include $500$ unoccupied states together with the six doubly-occupied states. The Formaldehyde molecule is oriented in the simulation in $x-y$ direction and the $\text{CO}$ stretching mode corresponds to a motion in the $x$-direction.

In all results, we calibrate the zero-point energy as the lowest ground-state energy for all nuclear configurations, i.e. the minimum of the ground-state potential-energy surface along the $\text{CO}$ stretch coordinate. To obtain the excited-state energies, we add the excitation energies obtained from Eq.~\ref{eq:matrix2} to the ground-state energies for each nuclear configuration.

\begin{figure}[t]
{\includegraphics[width=0.5\textwidth]{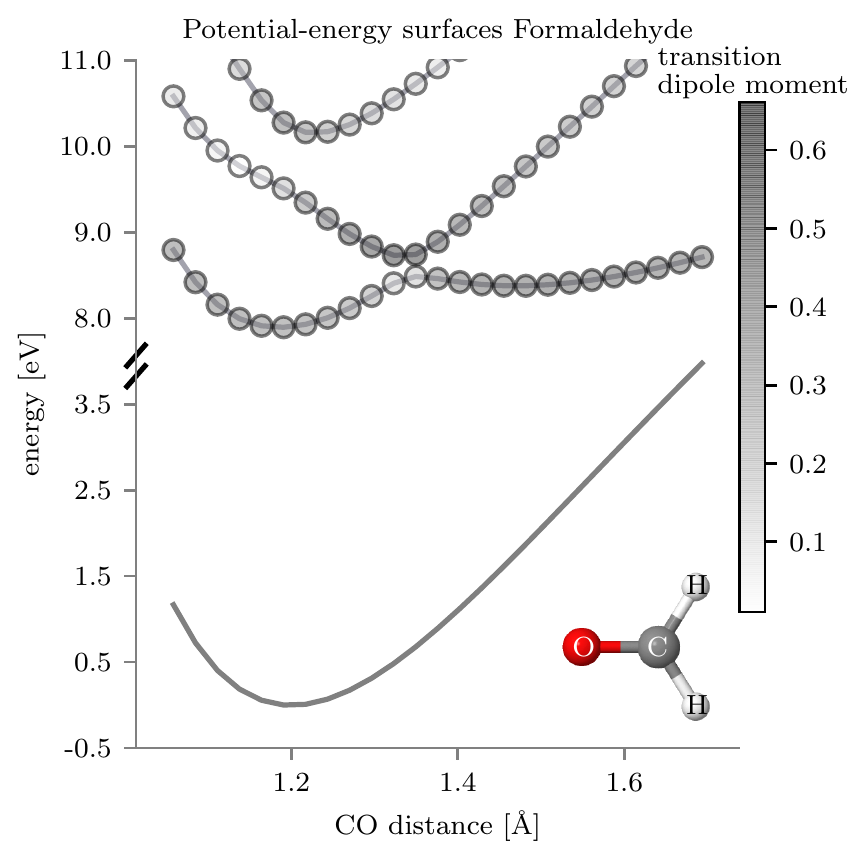}}
\caption{Potential-energy surfaces of the $^1\text{A}_1$ excited-state manifold of a Formaldehyde molecule along the $\text{CO}$ stretching mode. The lowest surface shows the ground-state potential-energy surface. The excited state potential-energy surfaces between $8$~eV and $11$~eV are color coded by their transition dipole moment in the $x$-direction out of the ground-state.}
\label{fig:01}
\end{figure}

\begin{figure*}[t]
{\includegraphics[width=\textwidth]{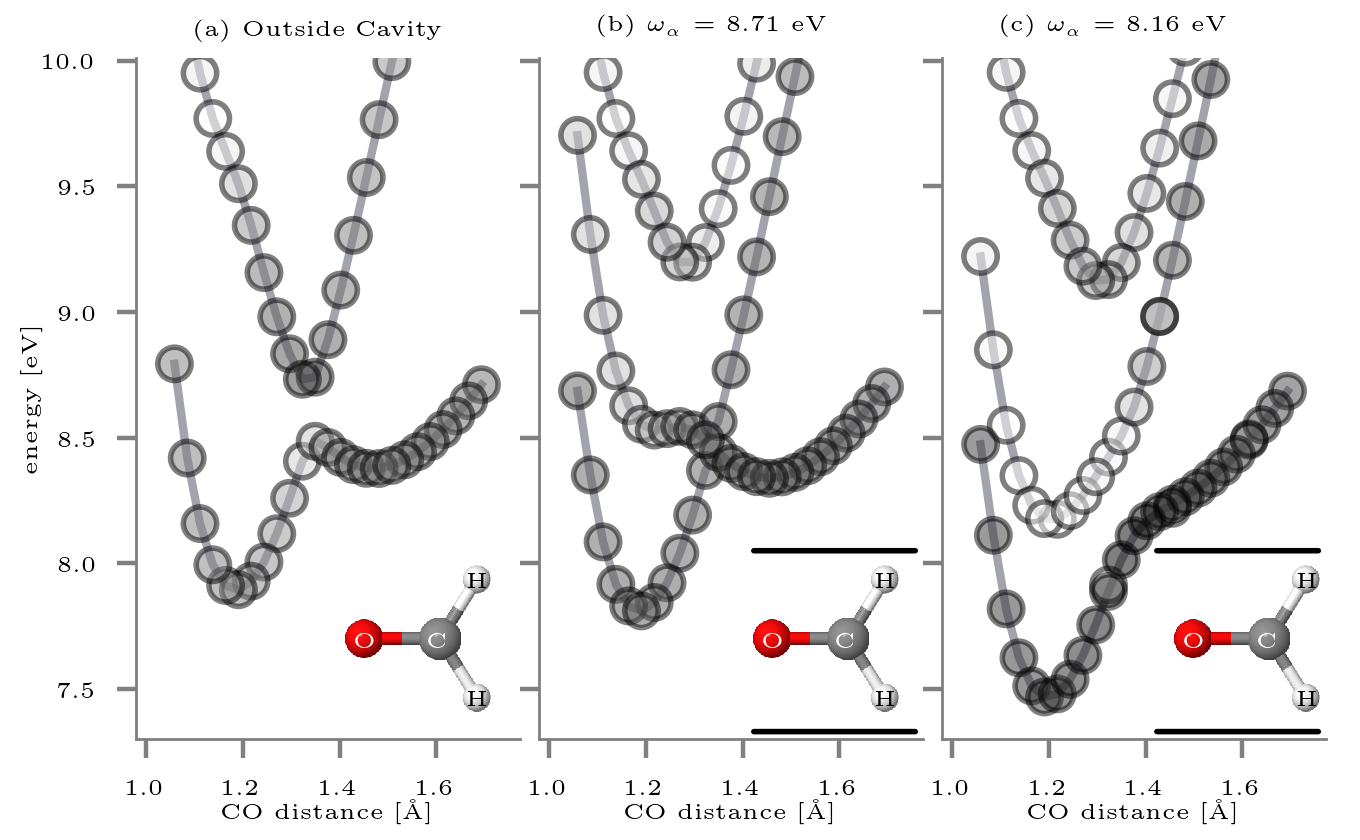}}
\caption{(a) A detailed view of the potential-energy surfaces of the $^1$A$_1$ excited-state manifold of a Formaldehyde molecule along the $\text{CO}$ stretching mode. The first and second excited-state surfaces exhibit an avoided crossing and a double well minima. (b) Formaldehyde molecule strongly coupled to a cavity mode of $8.71$~eV. The additional degree of freedom of the photon mode introduces an additional surface that quenches the avoided crossing. (c) Formaldehyde molecule strongly coupled to a cavity mode of $8.16$~eV. An additional polaritonic surface inhibits the right minimum at $1.48$~\AA ~and only the minimum at $1.21$~\AA ~remains.}
\label{fig:02}
\end{figure*}

First we construct the potential-energy surfaces for the Formaldehyde molecule along the $\text{CO}$ stretching mode without a coupling to an optical cavity. In Fig.~\ref{fig:01}, we show the ground-state potential-energy surface, as well as the excited-state manifold from $0$~eV to $11$~eV of the $^1\text{A}_1$ manifold of a Formaldehyde molecule. The lowest surface shown in grey depicts the ground-state potential-energy surface with a harmonic shape. In contrast, the excited-state surface shows a double well structure, with two minima at $x=1.2$~\AA ~and $x=1.48$~\AA. ~Between these two minima, the manifold exhibits an avoided crossing around $1.35$~\AA ~with an energy gap of $\Delta=0.4$~eV. The minimum at $x=1.2$~\AA ~is slightly shifted from the minimum of the ground state surface. The minimum at $x=1.48$~\AA ~is higher by $0.5$~eV then the left minimum. Additionally, the two higher surfaces shown in the figure also show an avoided crossing around $x=1.2$~\AA.

The excited-state potential-energy surfaces shown in Fig.~\ref{fig:01} between $8$~eV and $11$~eV are color coded. The color coding refers to the strength of their transition dipole moment from the ground state in the $x$-direction. For light-matter coupling, the transition dipole moment is an important quantity, since it mediates the light-matter coupling. We find that the transition dipole moment is highest at the maximum and minimum positions of the respective surfaces.

Next we discuss the excited-state manifold in more detail. In Fig.~\ref{fig:02} (a), we look at the excited state manifold between $7.5$~eV and $10$~eV. We see that the $\pi$-$\pi^*$ transition on the right has a stronger dipole moment. Also the upper surface has at the position of the avoided crossing a strong transition dipole moment. At around $9.5$~eV, we also see that the dipole amplitude changes due to the avoided crossing as discussed before. The state to the right is due to a $(\pi,\pi^*)$ transition, while on the left the excitation is of $(n,3p_y)$ nature~\cite{hachey1995, casida1996}. We find the higher minima at the right has a higher transition dipole element.


Fig.~\ref{fig:02} (b) shows then the Formaldehyde molecule strongly coupled to a cavity mode of $8.71$~eV. As ${\boldsymbol\lambda_\alpha}$, we use a value of 0.04. The effect of the cavity becomes clearly visible in the figure. Due to the additional degree of freedom of the cavity mode that is now strongly coupled, we find an additional potential-energy surface. The presence of the additional potential-energy surface pushes the upper surface up. Additionally the additional surface hybridizes with the $\pi-\pi^*$ transition leading to a closing of the avoided crossing, effectively quenching the avoided crossing. Fig.~\ref{fig:02} (c) shows a different setup, again the Formaldehyde molecule is strongly coupled to a cavity, in this case to a mode of $8.16$~eV with $\boldsymbol\lambda_\alpha=0.06$. The additional polaritonic surface inhibits the right minimum at $1.48$~\AA ~and only the minimum at $1.21$~\AA ~remains. While in Fig.~\ref{fig:02} (a) and (b), we find a double well structure with two minima for the lower excited-state surface, in Fig.~\ref{fig:02} (c) the lower surface has only one minimum.

\begin{figure*}[t]
{\includegraphics[width=\textwidth]{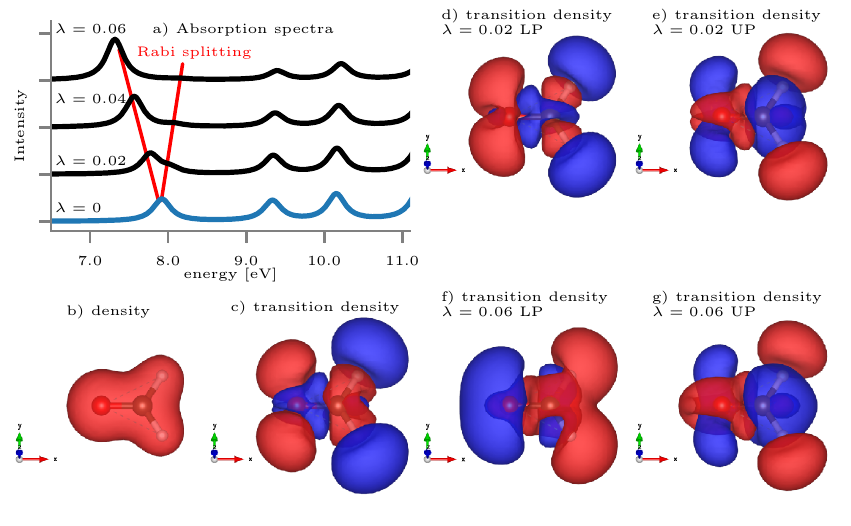}}
\caption{(a) Optical absorption spectrum of the Formaldehyde molecule for a $\text{CO}$ distance of $1.22$~\AA~ and optical cavity frequency of $\omega_\alpha=7.92$~eV. Increasing the electron-photon coupling strength $\lambda_\alpha$ introduces an redistributes spectral weight and introduces an additional peak and the Rabi-splitting. (b) Electron ground-state density for the parameters as in (a). We show the transition density for CO=$1.22$~\AA ~outside the cavity of the first peak in (c). In (d)-(e), we show the lower (LP) and upper polariton (UP) transition density for $\lambda_\alpha=0.02$, respectively and in (f)-(g) the same for $\lambda_\alpha=0.06$.}
\label{fig:03}
\end{figure*}

In Fig.~\ref{fig:03}, we show the absorption spectra and transition densities, as obtained from the QEDFT linear-response formulation for different electron-photon coupling strength. In Fig.~\ref{fig:03} (a), we show the absorption spectrum for a $\text{CO}$ distance of $1.22$~\AA. The blue curve shows the spectrum outside the cavity. Showing agreement with Fig.~\ref{fig:01}, we find three peaks in the range between $7-11$ eV. At this $\text{CO}$ distance, the ground state density is very homogeneously spread over the H$_2$CO molecule, as depicted in Fig.~\ref{fig:03} (b), since this state is close to the ground-state nuclear configuration. As a next step, we analyze the electronic transition densities that can be obtained by the linear-response formalism. The transition density is defined by:
\begin{widetext}
\begin{align}
\rho_{0i}(\textbf{r}) = \int dq_1...\int dq_{N_{p}} \int d\textbf{r}_2... \int d\textbf{r}_{N_e} \Psi^*_0( \{\textbf{r}_i\}, \{q_\alpha\})\Psi_i(\{\textbf{r}_i\},\{q_\alpha\})
\label{eq:transition_density}
\end{align}
\end{widetext}
where the integration runs over all $N_p$ photon modes, and all $N_e$ electronic variables except one. In Eq.~\ref{eq:transition_density}, $\Psi^*_0(\{\textbf{r}_i\},\{ q_\alpha\})$ is the electron-photon ground-state, and $\Psi^*_i(\{\textbf{r}\},\{ q_\alpha\})$ an electron-photon excited-state. The transition dipole moments that are also plotted in Fig.~\ref{fig:01} and Fig.~\ref{fig:02} are connected to the transition densities by $\textbf{d}_{0i}=\int d\textbf{r}\,\textbf{r}\rho_{0i}(\textbf{r})$.
The transition densities are shown in Fig.~\ref{fig:03} (c)-(g). In Fig.~\ref{fig:03} (c)-(g), the color coding is defined such that red regions imply density accumulation, while in blue regions the electron density is reduced by the electronic excitation. We note that there is a phase degree of freedom, so whether red is accumulation or depletion is an arbitrary choice. In a time-dependent picture, the transition density refers to the oscillating charge that is connected to the specific transition. The transition density of the first transition that is shown in Fig.~\ref{fig:03} (a) for $\boldsymbol\lambda_\alpha=0$, i.e. outside the optical cavity is shown in Fig.~\ref{fig:03} (c). This particular excitation as a transition density structure such that regions more far away from the molecule have opposite sign, and the same for regions close to the nuclei. We highlight the different sign between the CH$_2$ subgroup and the oxygen atom. 

\begin{table}[t]
\begin{center}
\begin{tabular}{ |c | c | c  c| c  c|}
\hline
 &  $\lambda_\alpha=0$ & $\lambda_\alpha=0.02$ (LP)&  (UP)& $\lambda_\alpha=0.06$ (LP)&  (UP)\\\hline
    ($n$,$3p_y$) & 99.34  & 62.13 & 77.43 & 48.01 & 86.32\\\hline
    ($\pi$,$\pi^*$) & 7.27  & 2.61 & 12.42 & 14.08 & 21.82\\\hline
    ($0$,$1$) & 0  & 78.13 & 61.40 & 85.99 & 43.65\\\hline
\end{tabular}
\end{center}
\caption{Amplitudes of Kohn-Sham transitions of the excitations as described in Fig.~\ref{fig:03} and the values are given in percentages.}
\label{tab:casida}
\end{table}

The spectral weight in the absorption spectra shown in Fig.~\ref{fig:03} (a) is conserved due to the sum rule of the total spectral weight~\cite{flick2018}. The sum rule specifies that the total spectral weight has to be equal to the number of electrons in the system. Since the number of electrons does not change, and the total spectral weight is independent of the photonic degrees of freedom, electron-photon interactions do not change the total spectral weight. As a consequence, strong light-matter coupling changes the relative spectral weight of peaks. If we strongly couple the system to a single photon mode, an additional pole will be predicted by our linear-response formulation. This is seen in Fig.~\ref{fig:03} (a), where the spectral weight of the lowest peak is redistributed between the lower (LP) and upper polariton (UP) by coupling strongly to a resonant cavity mode. In Fig.~\ref{fig:03} (d)-(g), we now show how this transition density is altered under strong-light matter coupling. Fig.~\ref{fig:03} (d) shows the LP transition density for $\lambda_\alpha=0.02$, and Fig.~\ref{fig:03} (e) shows the UP transition density. As seen in Fig.~\ref{fig:03} (a), the spectral weight of these two transitions is not identical, and also the transition densities are qualitatively different. The most prominent difference between the two densities is at the regions close to the Hydrogen atoms, where the LP has an additional phase change, and the areas around the oxygen atom. 

Differences between the LP and UP become even more pronounced if the light-matter coupling strength is increased to $\lambda_\alpha=0.06$, as shown in Fig.~\ref{fig:03} (f) and (g) for the UP and LP, respectively. Here, the UP has an additional phase change at the oxygen atom, and the area of the CH$_2$ subgroup differs significantly. For this coupling strength, the UP has a higher spectral weight then the LP. These changes in the transition density can also be analyzed from different mixing of Kohn-Sham transitions as shown in Tab.~\ref{tab:casida}. For the electronic transitions, we use the same name conventions as in Ref.~\citenum{casida1998}. While the transition outside the cavity mainly is of ($n$,$3p_y$) nature, under strong light-matter coupling also fractions of the ($\pi$,$\pi^*$) transition are mixed in. This different transition then leads to changes in the real-space picture of the transition density. We also note that this additional nontrivial mixing goes beyond a simple two-level description of the electronic manifold. In summary, since the transition densities change differently in the strong coupling, and not only the transition dipole moments, we expect also that other electronic observables are affected.

To conclude, in this work we present the first method to calculate polaritonic potential-energy surfaces from first principles. With a focus on the Formaldehyde molecule we show that strong coupling to a cavity mode can alter the potential-energy surfaces in various ways depending on the coupling-strength and the cavity frequency. These changes affect not only spectral weight, but will affect any physical observable, due to changes in the electron transition density.

Looking forward in future work, we suggest to connect to the studies in Refs.~\citenum{hachey1995, hachey1996, gomez2010} that are able to more accurately consider correlation effects by using the multireference configuration interaction method, and coupled cluster methods. These studies predicted that the second minimum in the lower surfaces at $x=1.48$\AA~ shown in Fig.~\ref{fig:02} is at lower energy, and either equal, or lower as the left minimum leading to a competing relaxation pathways. Quenching one of these minima by strong-light matter coupling has strong implications on the relaxation pathways. Additionally also effects of all nuclear degrees of freedom of the molecule, e.g. the $\text{HCH}$ bending mode, may have to be considered to get a comprehensive picture of the conical intersections~\cite{gomez2010} in the Formaldehyde molecule. Furthermore, the use of more accurate electron-photon functionals that can also capture two-, three-, many photon effects may be necessary to study strong light-matter coupling strength.

In summary, this work opens the possibility to study the modification of  photochemical reactions under strong light-matter coupling from first principles. Potential applications of strong-light matter coupling could be to either directly enhance the efficiency of the reaction, or to quench undesired chemical pathways with the possibility to controllably alter and design potential-energy surfaces. More specifically, this proof-of-concept calculation of the Formaldehyde molecule shows how strong coupling can be exploited to alter photochemical reaction pathways by influencing avoided crossings. In introducing an \emph{ab initio} method to calculate excited-state potential-energy surfaces, our work opens a new avenue for the field of excited-state nanophotonic and polaritonic chemistry.\\

\textbf{Acknowledgments:} This work was supported by the DOE Photonics at Thermodynamic Limits Energy Frontier Research Center under Grant No. DE-SC0019140. This work used resources of the National Energy Research Scientific Computing Center, a DOE Office of Science User Facility, as well as resources at the Research Computing Group at Harvard University. We acknowledge Andrew Winnicki for useful discussions, and Sebastian Ohlmann from MPCDF Munich for help with the parallel implementation. J.F. acknowledges fellowship support from the Deutsche Forschungsgemeinschaft (DFG Forschungsstipendium FL997/1-1). P.N. is a Moore Inventor Fellow and a CIFAR Azrieli Global Scholar.

\bibliography{light_matter_coupling} 

\end{document}